# Evidence of a cyclonic regime in a precessing cylindrical container

W.Mouhali, T.Lehner, J.Léorat, R.Vitry

LUTH, CNRS/UMR8102, Observatoire de Paris-Meudon, 5, Place J. Janssen, 92195 Meudon, France

*Abstract* : We report experimental observations obtained by particle image velocimetry (PIV) of the behavior of a flow driven by rotation and precession of a cylindrical container. Various hydrodynamical regimes are identified according to the value of the control parameter which is the ratio $\varepsilon$ of the precession frequency to the rotation frequency. In particular when $\varepsilon$ is increased from small values, we have observed an induced differential rotation followed by the apparition of permanent cyclonic vortices.

## 1- Introduction

Experiments with flows driven in a precessing container show an interesting variety of regimes according to the precession parameters. Different container geometries have been used, such as spheroids (Malkus 1968, Vanyo 1995, Noir et al 2001) and sphere (Goto 2007). Precession driving within a cylindrical container has been also the subject of numerous experimental studies: see for instance, Gans (1970a), Vladimirov& Tarrassov (1984), Mannasseh (1992,1994,1996), Kobine (1995), Meunier et al (2008) and Lehner et al (2010).

Our experimental configuration with a cylindrical vessel is motivated by fondamental physics but also by the fluid dynamo context, since it has been proposed that precession could be a good candidate for driving dynamo action in a conducting fluid (Malkus 1968, Gans 1970b). Dynamo action in a precessing container has indeed been numerically demonstrated rather recently using three different geometries : spherical (Tilgner 2005), spheroïdal (Wu & Roberts 2009) and cylindrical (Nore et al 2011).

Let us recall some specific features of rotating fluids relevant for precessing flows. As forcing increases from zero, the transition from solid body rotation to a laminar steady regime and a turbulent one is acknowledged and related to the dynamics of inertial waves. These waves were analytically predicted by Lord Kelvin (1880) as inviscid inertial modes in a rotating cylinder, were studied further on by Kudlick (1966) in the viscous case.

In a non ideal fluid, inertial waves are dissipated under viscosity and they can be excited through precession, which induce a forcing on the azimutal number m=1. The coupling of inertial waves inbetween them is also possible, for example the resonant coupling generating a third eigenmode can lead to parametric instability when the difference in the azimuthal wavenumbers is Δm=1 (Kerswell 1993, Mahalov 1993). Other instabilities such as the elliptical instability (for Δm=2) may also occur and have been also studied in details (see a review by Kerswell (2002)). In particular the coupling of an axisymmetric mode (m=0) with the forced helical mode with m=1 can lead to the generation of geostrophic flow in the case of precession with subsequent secondary instabilities (Lehner et al 2010,). This latter result does not contradict the often quoted paper of Greenspan (1969) which states that the self-coupling of a single mode to generate geostrophic flow is not possible in general since here we have a nonlinear interaction between *different* modes. A single mode can however couple with itself non linearly into a boundary layer including viscous effects and generates by reflection a standing wave (see for example Lagrange et al (2008)). This non linear modes mode coupling resonance due to the non linear advection term in the Navier-Stokes equation should be distinguished from the linear resonant cavity effect when a wavelength of an inertial eigen mode matches with the height of the cylinder.

The present work describes the occurrence of a new intermediate regime, between laminar et turbulent states, where a few isolated cyclones form and appear to be advected by the zonal azimuthal flow.

This paper is organized as follows : in section 2 , we describe our experimental set up with the PIV detection system. In section 3 we report our observations in the different regimes according to the value of the ε parameter, while section 4 proposes an outline and a conclusion.

## 2-Experimental set up and instrumentation

A right PMMA (plexiglass) cylinder of radius R= 145 mm and adjustable length L (280<L<450 mm) is filled with water and it is put into rotation at the angular velocity Ωo around its symmetry axis (defining the z direction). It is mounted on an horizontal platform rotating at the angular velocity Ωp. The cylinder spin axis is tilted relative to the rotation axis of the platform with a fixed angle of 90°. Each rotation axis has its own driving motor, so that the angular velocities can be varied independently, there are limited to 60 rad/s for Ωo and to 6 rad/s for Ωp. The aspect ratio h=L/ 2R can be varied between about 1 and 4/3 with an internal lid acting as a piston. The fluid viscosity is kept fixed , since water remains at laboratory room temperature. The two experimental control parameters are the aspect ratio h and the precessing parameter ε=(Ωp/Ωo) << 1. The whole experimental setup is sketched on figure 1.

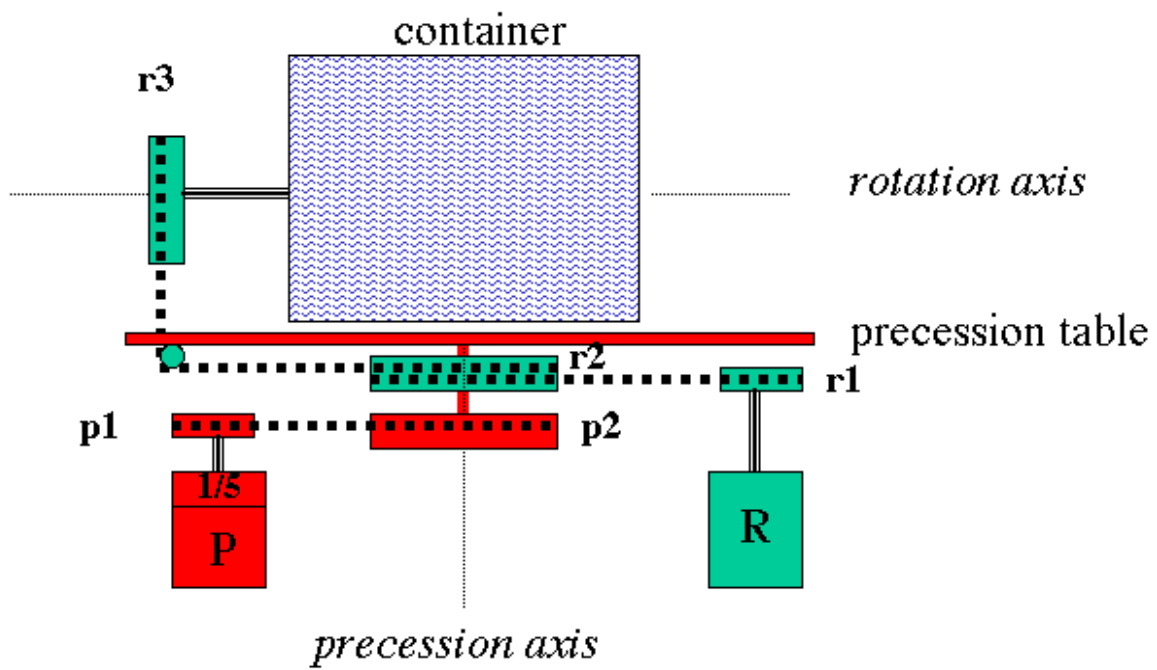

**Fig1** Sketch of the cylindrical container with its rotation and precession axes (r1,r2, r3 and p1,p2 refer to transmission from rotation and precession drives (R and P)

The principles of the PIV method are well known and we need to present here only some peculiar features of the instrumentation.

For the PIV measurement, a JAI M1 CCD camera (with an acquisition speed of 12 images/s, with a rectangular field of 1280*1024 pixels) is corotating with the container with its optical axis along the rotation axis z. It is mounted flush on one circular window in the center of one lid of the cylinder. This optical configuration allows then to see directly differential rotation effects. The lightning of the measurement plane is provided through two synchroneous flashes delivered by two independent Xenon light lamps (with a controlled time delay between two flashes). The light plane thickness is typically about 10 mm. The fluid is seeded with small reflecting particles (density 1.03g/cm3, diameter 250 µm). Their relatively large diameter is adapted to the amount of light power available. The recorded pairs of images of the PIV are transferred and stored on a PC for later analysis. Note that since the camera frequency is fixed, the time delay between two snapshots of a pair is 2*1/12=1/6 s (167 ms) . This sampling timis much greater than the typical non linear time of small scales (for example L/v= 10 ms if L=1 cm and the relative flow speed v is v= 10 cm/s) and is thus not convenient to study the turbulent regime. Another limitation concerns the field of view : since the distance between the measurement plane and the camera, which is corotating with the container, is of the order of the radius , the field of view does not generally include a full section of the flow, except when the light plane is pushed towards the container end.

By comparing two consecutive pictures using a DPIV software due to Meunier & Leweke (2003), a 2D velocity flow noted as (u,v) is obtained on a regular 60 by 60 cartesian grid,. The spatial resolution (in pixel/cm) is fixed according to the camera aperture and to the location of the light plane such that the flow velocity v is known in cm/s. The axial vorticity (in rd/s) is also simultaneously obtained (on a similar grid). For a rotation rate of one turn/s which corresponds to a wall speed close to 1 m/s, the «large scale» Reynolds number is Re = Vo R/ν ≈1.32 $10^5$, for ν = $10^{-6}$ m2/s, R = 0.145 m and the Ekman number is defined as Ek=1/Re.

In order to complete the numerical treatment, the DPIV cartesian velocity field is conveniently expressed in polar coordinates centered on the cylinder axis, which coincide with the optical axis of the CCD camera. The interpolation on a polar grid with the selected resolution is performed using the Matlab software and yields the transverse velocity field (uradial, uazimutal) or ($u_r$, $u_\phi$).

Alternatively, the light plane may also be chosen to be a meridian plane, and the optical axis of the camera is then directed radially parallel to the precession axis and fixed in the precession frame. In principle, this setting should be convenient to observe the axial and radial velocities and the azimuthal component of the vorticity. In practise, since the flow consists mainly of an azimuthal advection by the rotation, which has a null projection on the light plane, we have found that this configuration could not give significant information on the axial circulation.

The experimental protocole is the following : first we put the cylinder into rotation to a selected angular velocity and wait to reach the stationary regime of rigid rotation and then we put on the precession. In the two cases we can choose a given linear time sweeping ramp (typical values are $d\Omega_0/dt=0.1$ tr/m/s and $d\Omega_p/dt=0.02$ tr/m/s).

**3- Experimental results**

We describe now our observations. According to the value of the control parameter $\varepsilon$ we have identified several different regimes for the fluid behavior at a given aspect ratio h (ratio of length L to diameter 2R). As recalled above a linear cavity mode resonance may occur when the axial wavelength of a Kelvin mode coincides with the length of the cylinder, this is for example the case if h is close to 0.98. Although the resonance cavity lengths form a dense set, damping by viscous effects is efficient, except for a few set of aspect ratio. We choose here h = 1.17 as a typical non resonant configuration.

a) small precession rates : laminar steady flow

For very small $\varepsilon$ ranging up to to about $10^{-2}$, the flow appears as a quasi-steady solution in the precessing frame and vorticity distributions remains smooth: the flow behaves close to the stationary linearized solution which can be obtained by perturbation theory around the solid body rotation state of the Navier Stokes equation with $\varepsilon$ as a small forcing parameter. This steady solution in the precessing frame is a superposition of Kelvin modes m=1 satisfying the boundary conditions on the container walls (r=1 et z= ± a). One observes the predominance of the inertial mode with m=1 that is directly forced by precession. This may be shown qualitatively by using the buoyancy of air in the rotating flow to trace the line of minimal pressure: this form an S-shaped curve due to the axial circulation induced by the m=1 mode (see figure 2 below). This S shape can be predicted analytically within the linearized theory of the flow'description and accounting for the boundary condition on the two endwalls.

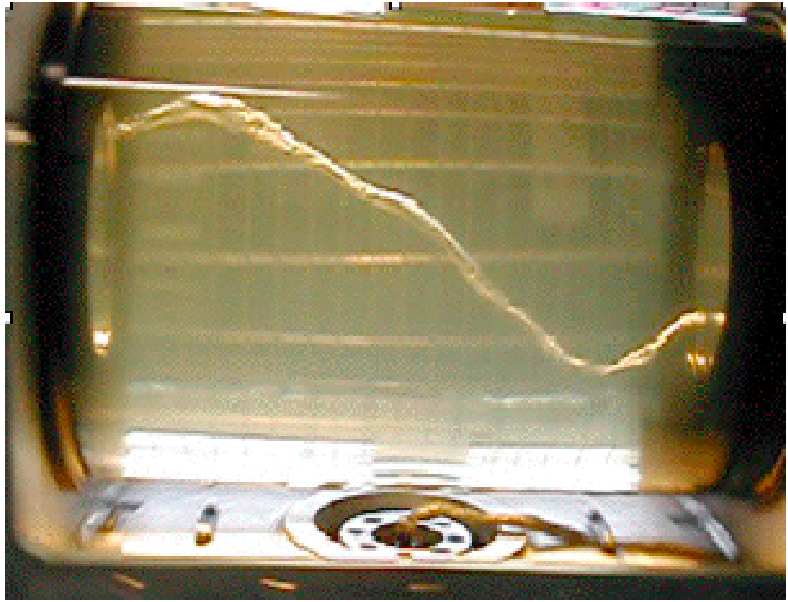

**Fig 2** Direct visualisation of the m=1 mode using buoyancy of air bubbles which follow the line of minimal pressure

This can also be shown more quantitatively using the PIV équipement. The most direct way to show the precession forcing of such an inertial wave on the azimuthal mode m=1 is to visualize the trajectories of advected particles acting as lagrangian tracers. This is easy to obtain by adding on a single image a few successive snapshots taken with a single flash for every frame (time interval is 1/12 s), which results in « particle tracking velocimetry».
 In the frame of the camera corotating with the container, the solid rotation (m=0) vanishes and the m=1 structure is a steady wave rotating with an angular velocity opposite to the one of the container, which induces on each fluid particle a rotating velocity, with a local amplitude fixed by by the wave radial profile and thus depending on the initial position of the tracer.

Figure 3 shows such lagrangian trajectories for ε =2.3 $10^{-3}$. Linear analysis show that typical wave speed is proportional to the product of the precession rate ε by the curved wall speed,. Figure 4 shows that indeed the diameter of the particle trajectories (in the central region) scales linearly with the forcing parameter ε in this weak regime.

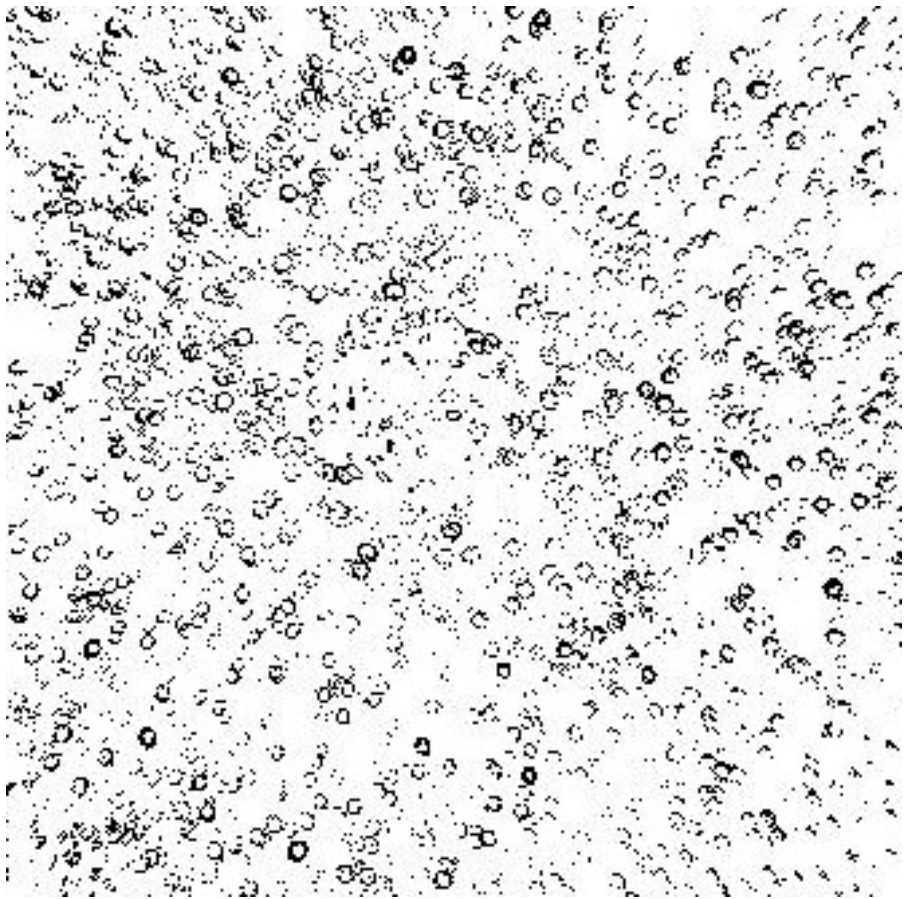

**Fig3** Lagrangian trajectories in the container frame (for h=1.17, Re=141 000) obtained by adding 12 consecutive snapshots

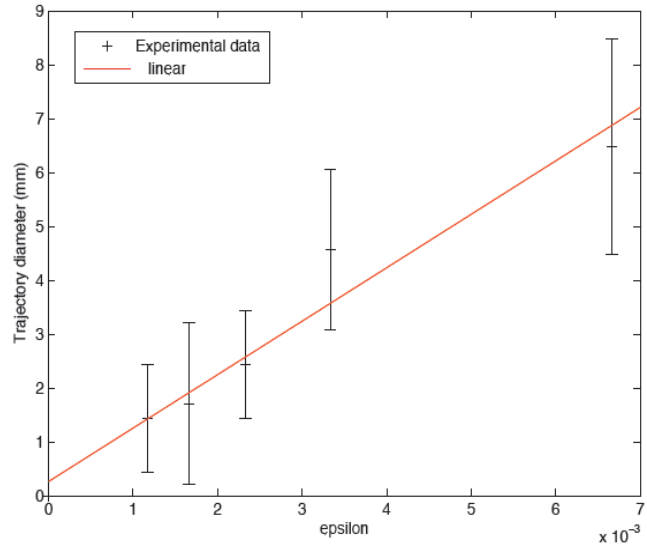

**Fig 4** Diameter of closed trajectories as a function of the precession rate ε (for h=1.17, Re=141 000).

As expected, the trajectories close after a rotation period and they are approximatively circular in the central region where the wave amplitude (a function of r and the azimutal angle) is relatively uniform, and become elliptical with a reduced amplitude when reaching the boundary where the inertial wave amplitude decreases and differential rotation is present. The amplitude of the m=1 mode may also be measured using PIV from the time history of the radial velocity in the center. For small precession rates, the central amplitude A is found to follow a linear law, A (mm) = 348 ε, so that a satisfying agreement is found between PTV (figures 4) and PIV (fig 5).We do not intend to pursue here more quantitative comparisons with the analytic steady solution since the linearized approximation is fully justified in this regime. Note that Meunier et al (2008) have studied **t**his regime in details using a configuration using a small precession angle.

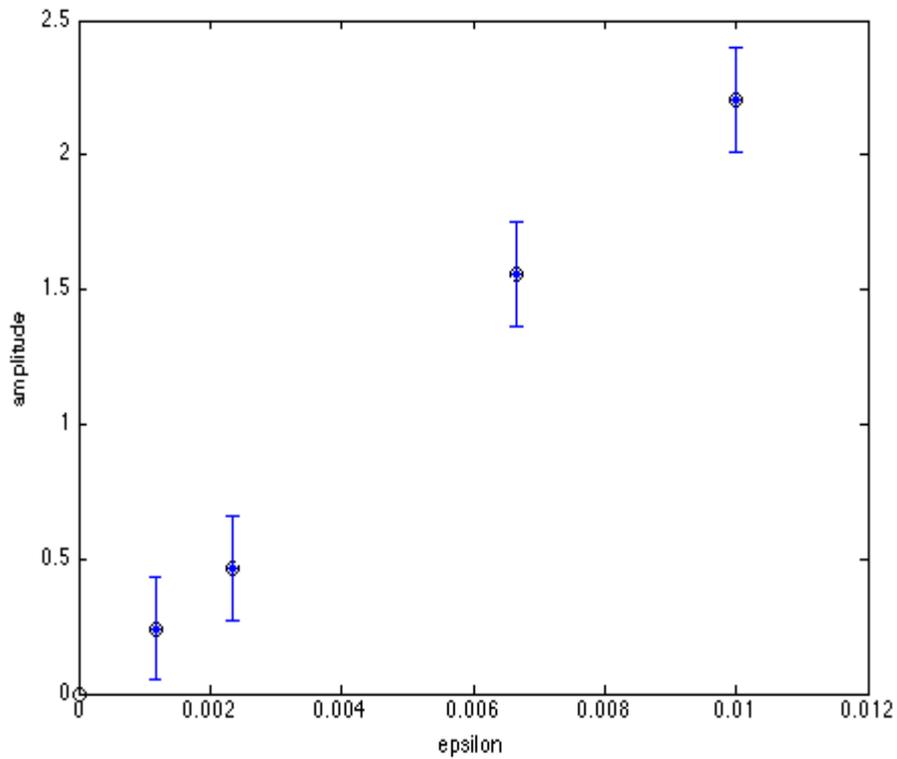

**Fig5** Amplitude of m=1 mode measured with PIV (same parameters as for PTV of figure 3)

For weak precession rates, the axial vorticity is found to remain quasi uniform in a domain near the center of the cylinder and the global rotation of the fluid is close to the solid body rotation at the imposed rotation frequency. By increasing the precession rate, the flow deviates from the linearized solution and differential rotation of the azimuthal flow sets in, see figure 6. This differential rotation can be described analytically using a weakly non linear approach (Lehner et al 2010).

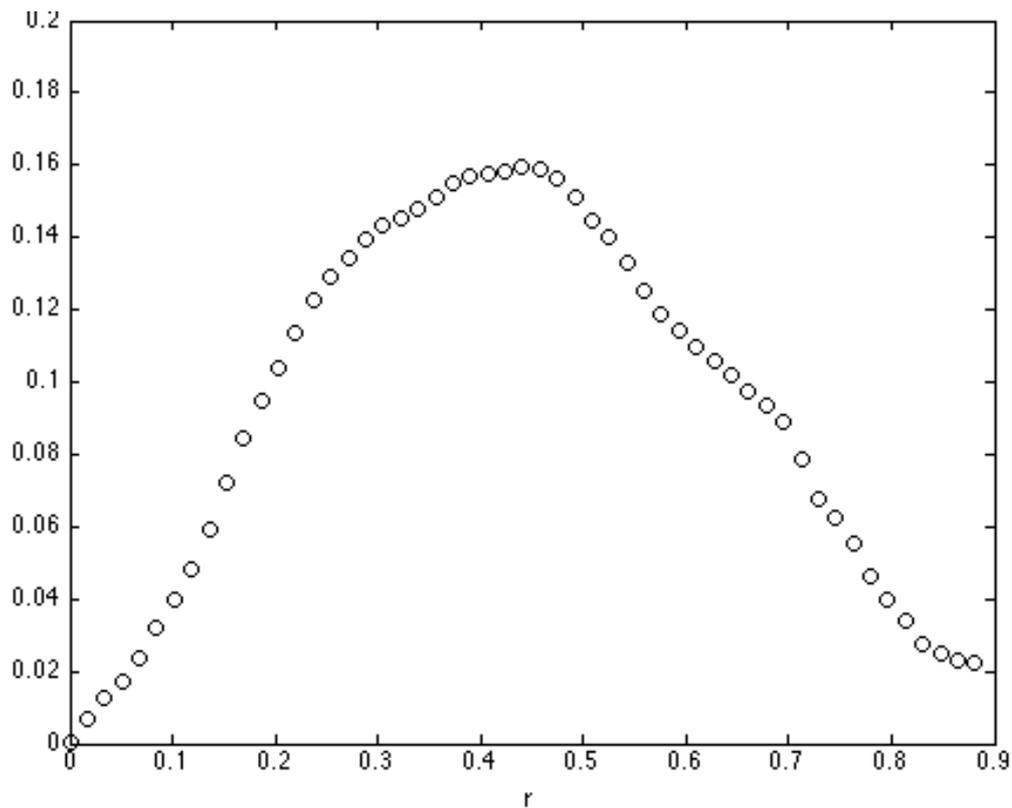

**Fig 6** Radial profile of the mean azimuthal velocity illustrating the differential rotation (for $\varepsilon$=??)

Note that the center of the container (r=0 and z=0), which is a symmetry center for the velocity boundary conditions and for the Coriolis acceleration (expressing precession forcing) is also a symmetry center of the linear solution. Parity breaking is expected when the fluctuation amplitude grows, but it cannot be visualized simply using a transverse section of the tank (the light plane of the PIV). Increasing the precession rate, the steady flow in the precessing frame indeed bifurcate to a new regime.

b) cyclonic regime
Keeping h=1.17 fixed, when $\varepsilon$ exceeds a critical value of about 0.02, the flow changes radically and a new flow regime sets in. In a cross section of the flow normal to the rotation axis, a few vortices become visible, rotating in the same direction as the one of the container, as may be seen on figure 7.

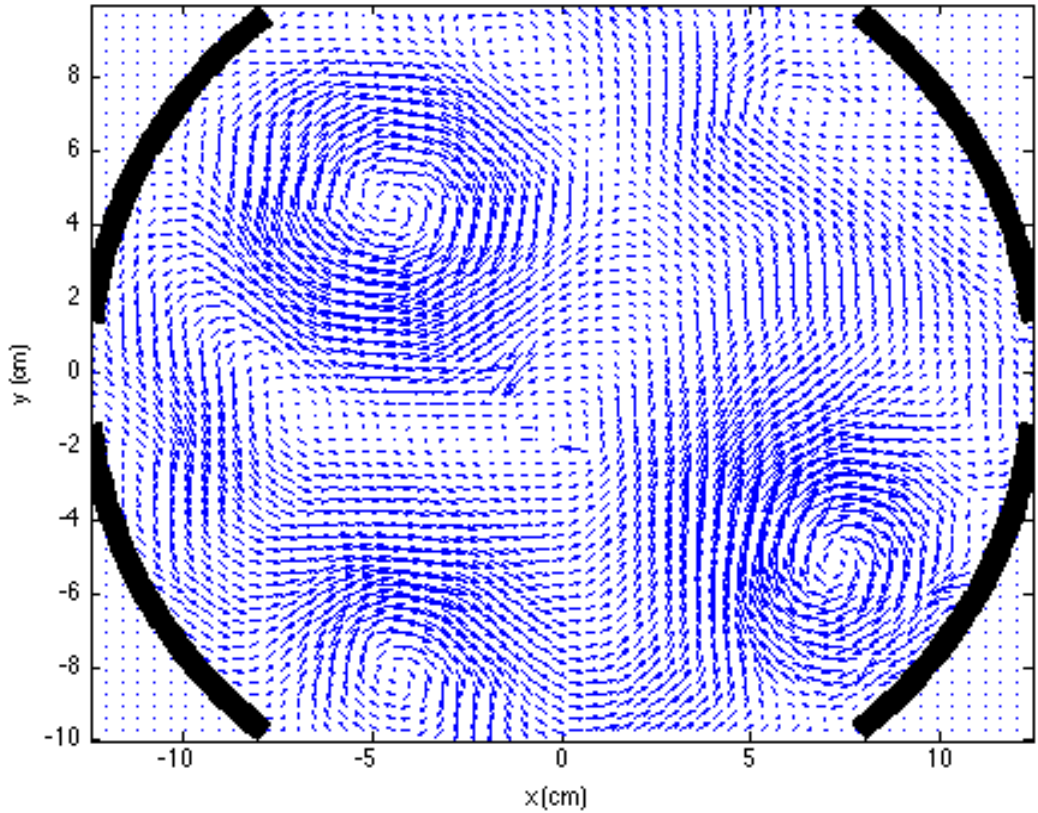

**Fig 7** Flow in the cyclonic regime : distribution of velocity in the x-y plane (for h=1.17, $\varepsilon$=0.033, Re=141 000)

The axial vorticity appears to become strongly concentrated in a few disconnected islands, and the relative vorticity maxima (about 8) are not far from the absolute vorticity of the bulk rotation ($4\pi$ for a spin frequency of 1 turn/s).

The flow in this cyclonic regime is not steady in the precession frame. Figure 8 shows six snapshots of the flow vorticity in the central region, separated by the container rotation period (1s). The time increases from top to bottom and left to right and note that the color table changes for each frame.

In this sequence, it appears that the number of cyclones remains unchanged but they are not fixed. Defining the cyclone centers as the points where vorticity is locally extremal (negative in the present case), one can follow the cyclones trajectories in the precession. Figure 9 shows a typical trajectory. The mean motion is a rotation around the cylinder axis, and it displays an advection of the cyclones by the differential rotation of the bulk flow. Superimposed oscillations of small amplitudes are also visible. Although they are clearly undersampled in the present case with 1/6 s between two consecutive positions, they show that the cyclones are also advected by the inertial waves in the container.

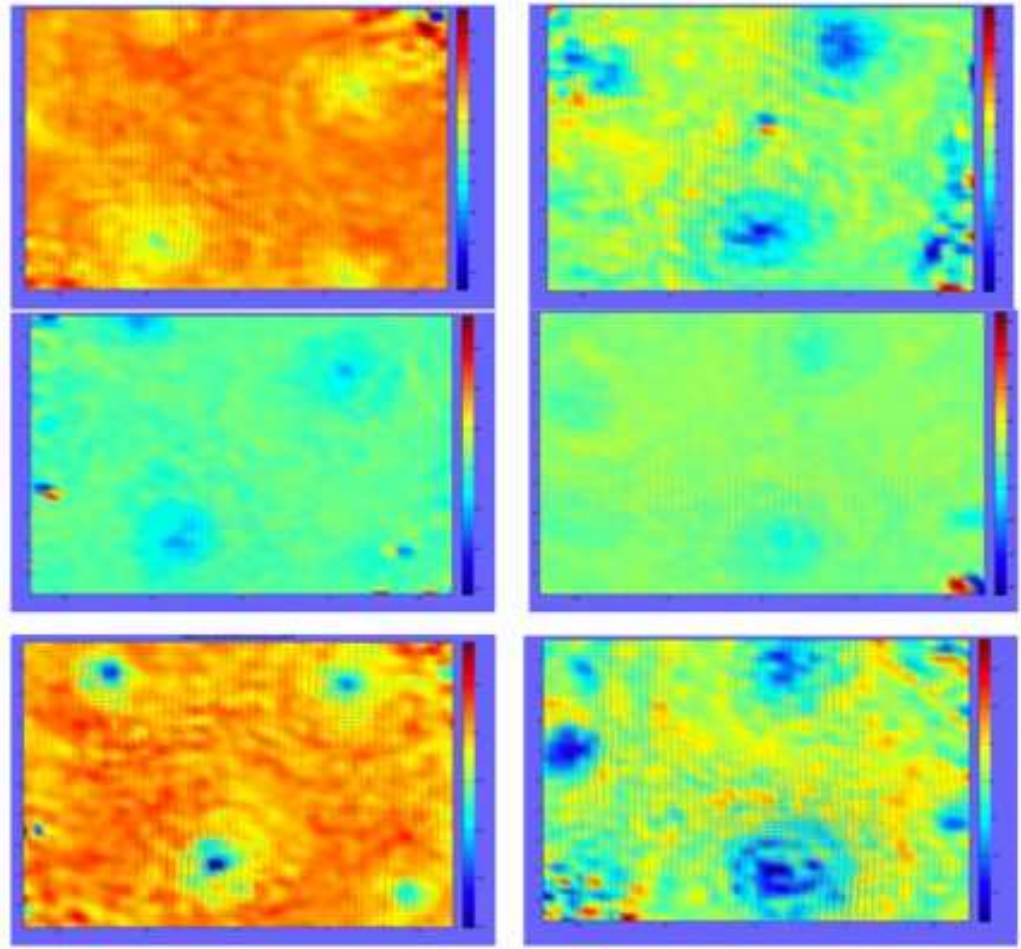

**Fig8** Successive vorticity fields in the cyclonic regime: time is running from top to bottom, first figure top left then down left and so on (for h=1.16, ε=0.033, Re=141 000, time interval is the rotation period =1s)

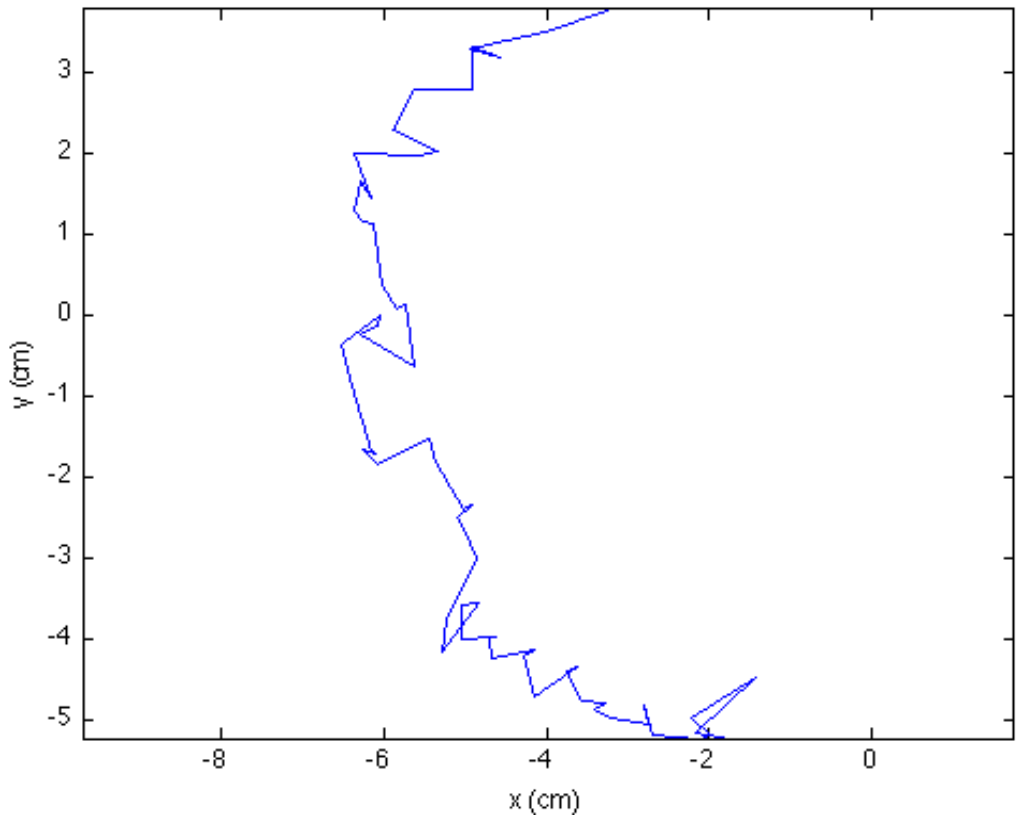

**Fig 9** Trajectory of a cyclone center in a transverse plane (for h=1.16, $\varepsilon$=0.033, Re=141 000)

The cyclones may be visualized through different transvers sections orthogonal to the rotation axis, with main characteristics which seem to be independent on the plane position. The observations suggests that these cyclones are elongated structures, parallel to the container axis, and advected by the differential rotation. These features presumably break the centro-symmetry of the flow, and the obvious issue to consider then concerns its axial structure. It would be interesting to confirm this hypothesis directly by complementary PIV measurements. We have tried to use a light sheet in a plane containing the rotation axis with a radially viewing camera. Since the camera is no more corotating with the container but is fixed in the precession frame, the cyclones are supposed to cross the light plane now and then. Moreover the flow induced by an axial cyclone crossing the light plane has no component in this plane. This is a penalizing geometry for a velocimetry measuring the transverse speed as is PIV, which may explain that we have not been able to confirm the axial structure of the cyclones with the available PIV facility. More involved methods, such as laser

Doppler velocimetry, measure a single flow component at one point and cannot help to describe the axial structure. Note also that these cyclones are not observed in direct numerical simulations, using the same geometrical and forcing parameters and this is plausibly a consequence of the low Reynolds numbers within computational reach ($Re<10^4$), compared to the experimental ones ($Re>10^5$). It could be also a consequence of a large angle of precession in cylindrical geometry, since they have not been observed either in the many experimental settings using spheroidal containers or in the case of small precession angle of Meunier et al (2008).

When the aspect ratio h varies, a cyclonic regime is always observed within a well defined range of precession rates. Note also that the bifurcation to the cyclonic regime occurs earlier when the aspect ratio of the container is close to resonance. Figure 10 shows such a case, with h=0.99, obtained for $\varepsilon = 0.01$, where one can also note that the cyclones may be not so well defined than in non resonant configurations.

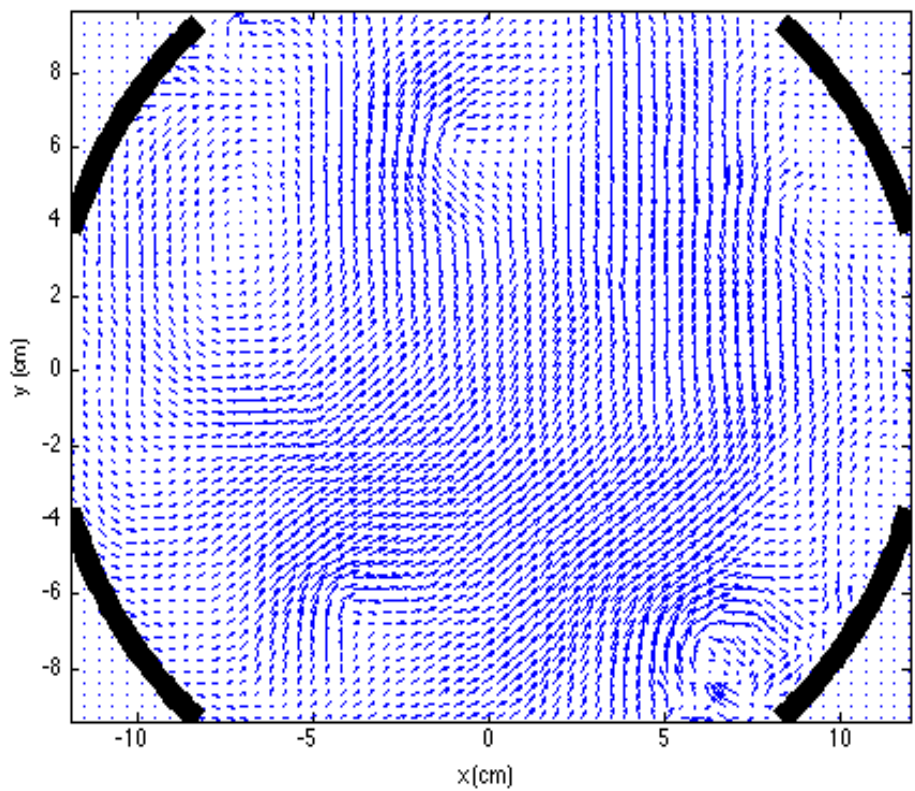

**Fig10** Cyclonic flow for the resonant cylinder (for h=0.99 ε=0.01, Re =141 000)

**Transition to the cyclonic regime**

The cyclonic regime is presumably a consequence of non linear couplings in the precessing cylinder when the inertial wave amplitude reaches a critical level, but the basis of its dynamics is not known. We recall that the origin of analogous vortices which are observed in some geophysical or astrophysical situations is still a matter of debate and they may be cyclonic or anticyclonic according to the physical settings.

In the present experimental approach, it may be useful to describe the initial steps of the instability which give birth to them. Starting with a precession rate below the transitional value, a diffuse layer of particles form on the curved wall. This accretion may be attributed to a centrifugation effect due to a small difference of density between particles and the fluid.

When the precession rate exceeds a critical value, wall particles are suddenly ejected radially towards the center, advected by a few vortices with a typical radius of the order of 1/3 of the container radius, see Figure 11.

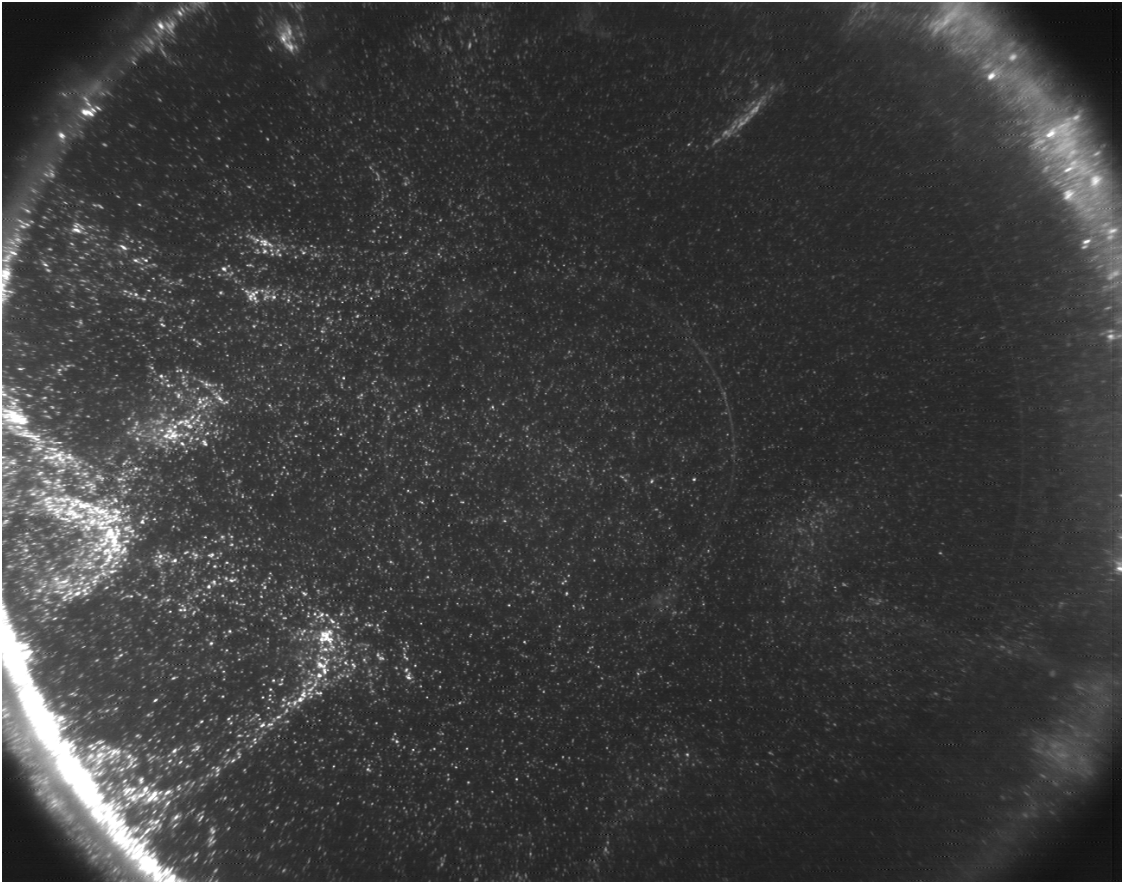

**Fig11** Transition from laminar to cyclonic regime wall : note the eruption of particles from the wall

c) Transition to turbulence and hysteresis

When ε increases, the differential rotation is enhanced (the inner part of the flow rotates slower than the wall) and above another critical precession rate ε, the cyclones are found to collapse into an unique vortex occupying the whole section of the cylinder and this unique vortice is found to rotate around a center close to the cylinder center of symmetry (see figure 12).

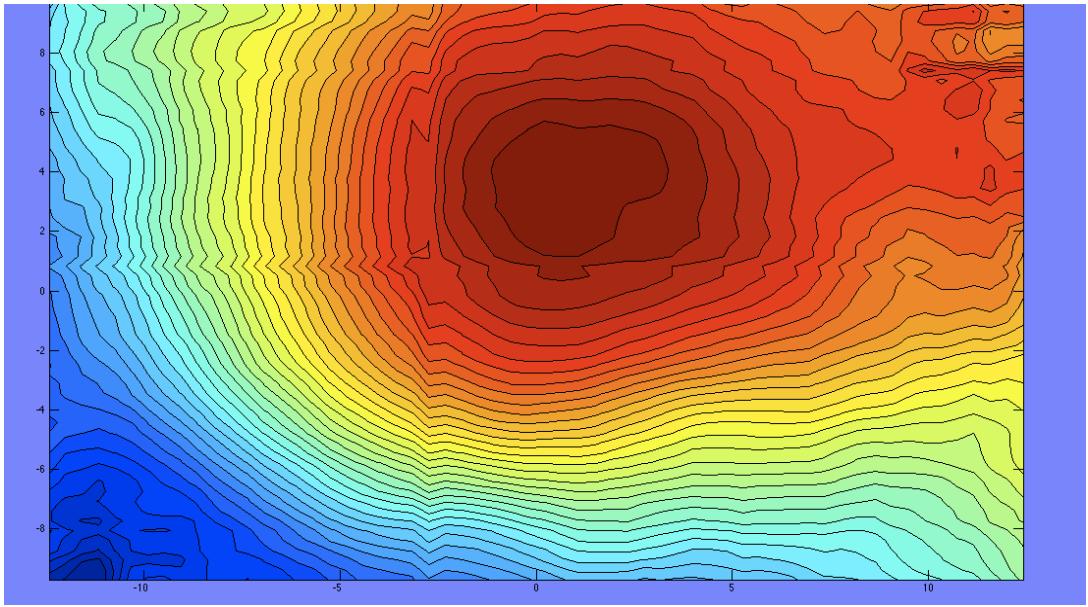

**Fig12** Typical single vortex flow between the regime with isolated strong cyclones regime and the fully turbulent regime ( h=1.17, Re= 141 000, eps=??)

The disappearance of the cyclonic regime may be attributed to the strong shearing action of the increasing differential rotation observed when the precession rate increases.

Increasing further the precession rate leads finally for $\varepsilon > \varepsilon_t^+$ (close to 0.1 for h=1.17*)* to a statistically steady turbulent regime, where the bulk rotation looks as vanishing and small scales are excited. The PIV is not able to give quantitative informations due to the relative flow speed, but we have verified that velocity measurements in the turbulent regime at one given point remain feasible using a LDV equipement put on the precessing table  In this regime, the global rotation is found to be inhibited and only small scales are excited (Y. Charles et al, in preparation).

This nonlinear regime verifies an hysteresis cycle: starting from a turbulent state, and decreasing the precession rate, the transition to the laminar regime occurs for another critical precession rate $\varepsilon_\tau^= < \varepsilon_t^+$ (The value of $\varepsilon_\tau^=$ depends of the experimental procedure used). This hysteresis cycle has already been described for the rotation torque when increasing or decreasing the precession rate (Gans 1970b).

**4-Outline and conclusion**

Precession driving deserves a practically interesting feature which is the ability to finely tune the forcing amplitude at will, at the largest available scale. Once the precession angle is fixed, as in the ATER facility, one is able to act on the spin and precession angular speeds and observe symmetry breakings of the flow. Starting from solid body rotation and increasing the precession rate from zero, a laminar steady 3D flow is observed for small precession rates, it becomes time dependent for larger precession rates, and increasing the precession rates (> 0.1, say) the resulting flow becomes fully turbulent and an hysteresis cycle is observed.

While PIV is well suited for the laminar regime, the sampling time of our camera (1/6 s for ATER experiment) is too low for suitable turbulence studies. However, when the precession rate lies in some intermediate range, we have shown above that using a camera corotating with the container, the PIV facility is able to detect a «cyclonic regime», which does not seem to have been observed previously. The observational results reported here are obtained at a rotation frequency of 1 turn/s, which correspond to a Reynolds number close to $1.32\ 10^5$. Direct numerical simulations using the same driving parameters (precession angle = 90°, precession rate = 0.033, aspect ratio L/2R= 1.17) and a rotation Reynolds number equal to 2 000 do not show cyclonic features (C. Nore, private communication). This gives a supplementary hint in favor of a nonlinear origin of this flow regime.

A theoretical analysis of this regime is not attempted here, but some basic remarks are in order. When the transition occurs, the differential rotation of the bulk flow is not small and is such that the flow close to the wall rotates faster that the central part. In absence of precession, this is a stable configuration for an axisymmetric pertubation in an ideal fluid according to the Rayleigh criterion. In a viscous precessing flow, a boundary layer forms, with a gradient (i.e. vorticity) which increases with the precession rate. The stability analysis of these mixed boundary layers is not available yet (in a purely rotating fluid, they are known as Stewartson layer on the curved wall and Ekman layer on the flat endwalls). They may become unstable and drive inertial waves of sufficiently high amplitudes to induce a transition to a new non linear regime where vorticity appears to be condensed in a few axial lines. An alternative approach in the framework of an inviscid flow is not excluded also (T.Lehner et al in preparation). Non axisymmetric modes may become unstable in a differentially rotating vortex according to the analysis of Billant et al (2005). Note however that the axial circulation due to precession and its induced anisotropy adds in all theoretical approaches a supplementary difficulty for the stability analysis.

*Acknowledgements*: we are gratefull to P. Meunier who has provided us his help with the DPIV software.